\documentclass[twocolumn,iop,numberedappendix,appendixfloats]{openjournal}
\usepackage[utf8]{inputenc}
\usepackage{amsmath,amssymb,amstext}
\usepackage{graphicx}
\graphicspath{ {./images/} }
\usepackage{natbib}
\usepackage{subcaption}
\usepackage[colorlinks,allcolors=blue]{hyperref}
\usepackage{dsfont}
\usepackage{xcolor}
\usepackage{listings}
\usepackage{fontawesome}

\newcommand{\daa}{\Delta\alpha/\alpha}

\begin{document}
\journalinfo{The Open Journal of Astrophysics}
\submitted{Submitted XXX; accepted YYY}

\shorttitle{Searching for new physics using high precision absorption spectroscopy}
\shortauthors{Webb et al.}
\title{Searching for new physics using high precision absorption spectroscopy; continuum placement uncertainties and $\daa$ towards the quasar PHL957}

\author{John K. Webb$^\dagger$}
\affiliation{Institute of Astronomy, University of Cambridge, Madingley Road, Cambridge, CB3 0HA, UK}
\affiliation{Clare Hall, University of Cambridge, Herschel Rd, Cambridge, CB3 9AL, UK}
\affiliation{Big Questions Institute, Level 4, 55 Holt St., Surry Hills, Sydney, NSW 2010, Australia}
\email{$^\dagger$jw978@cam.ac.uk}

\author{Chung-Chi Lee$^*$}
\affiliation{Big Questions Institute, Level 4, 55 Holt St., Surry Hills, Sydney, NSW 2010, Australia}
\email{$^*$lee.chungchi16@gmail.com}

\author{Dinko Milakovi{\'c}}
\affiliation{Institute for Fundamental Physics of the Universe, Via Beirut, 2, 34151 Trieste, Italy}
\affiliation{INAF - Osservatorio Astronomico di Trieste, via Tiepolo 11, 34131, Trieste, Italy}

\author{Darren Dougan}
\affiliation{Big Questions Institute, Level 4, 55 Holt St., Surry Hills, Sydney, NSW 2010, Australia}

\author{Vladimir A. Dzuba}
\affiliation{School of Physics, University of New South Wales, Sydney, NSW 2052, Australia}

\author{Victor V. Flambaum}
\affiliation{School of Physics, University of New South Wales, Sydney, NSW 2052, Australia\\}

\begin{abstract}
Detecting or placing upper limits on spacetime variations of fundamental constants requires quantifying every potential source of uncertainty. We continue our previous study into the impact of continuum variations on measurements of the fine structure constant, here in the context of quasar absorption systems. An automated (hence objective and reproducible) continuum modelling method is reported in an accompanying paper. We apply the method to the $z_{abs}=1.7975$ absorption system towards the quasar PHL957. Multiple continuum fits are generated, and for each, we derive independent models of the system, each giving its own measurement of the fine structure constant $\alpha$. This process isolates and quantifies the error contribution associated with continuum placement uncertainty. This source of uncertainty, ignored in many previous measurements, arises in two ways: (i) slight local continuum tilt uncertainty generates small line shifts, and (ii) different continuum estimates produce slightly different kinematic structures in the absorption system model. Taking continuum placement uncertainty into account, the new PHL957 measurement we obtain is $\daa = -0.53^{+5.45}_{-5.51} \times 10^{-6}$. This measurement assumes terrestrial magnesium isotopic abundances. Recommendations are provided for future $\alpha$ measurements. Finally, we also note the potential importance of the effects identified here for future redshift drift experiments.
\end{abstract}

\maketitle

\section{Introduction} \label{sec:Intro}

This paper is a companion paper to \cite{Lee2024}, in which we introduce an automated continuum fitting algorithm, based on cubic splines, designed for use on high resolution astronomical spectra. 

Modelling absorption features in high-resolution quasar spectra requires an estimate of the unabsorbed continuum level. Broadly, there are three options: (1) estimate $I_{0,\lambda}$ simultaneously with estimating the theoretical $\tau_{\lambda}$, (2) estimate $\tau_{\lambda}$ and $I_{0,\lambda}$ independently, or (3) make a preliminary estimate of $\tau_{\lambda}$ and $I_{0,\lambda}$ and subsequently simultaneously refine the parameters for $I_{0,\lambda}$ whilst minimising $\chi^2$ to model $\tau_{\lambda}$. (1) can be applied but only in limited circumstances: the continuum function must be simple, otherwise it is easy to create degeneracy between $I_{0,\lambda}$ and $\tau_{\lambda}$ unless the absorption feature is flanked by large continuum regions (which may not be the case). (2) can also be used but then the uncertainty estimates for the final absorption line parameters do not reflect continuum uncertainties. (3) alleviates the problems associated with the first two methods and is thus preferred.

For simplicity, consider only rest-frame quantities, i.e we ignore redshift in this illustrative discussion: let us take the true wavelength of an absorption line to be given by
\begin{equation}
    C = \sum_i^n\Delta\lambda_i\left( I_{0,i} - I_{obs,i}\right) \bigg/ \sum_i^n\left( I_{0,i} - I_{obs,i}\right), \label{eq:obscentroid}
\end{equation}
where $I_{0,i}$ is the true continuum level against which absorption takes place, $I_{obs,i}$ is the observed intensity, $\Delta\lambda_i$ is the pixel width of the $i^{th}$ pixel, and the summations are made over all $n$ pixels for the absorption line being measured. However, the measured wavelength is
\begin{equation}
    C' = \sum_i^n\Delta\lambda_i\left( I'_{0,i} - I_{obs,i}\right) \bigg/ \sum_i^n\left( I'_{0,i} - I_{obs,i}\right), \label{eq:truecentroid}
\end{equation}
where $I'_{0,i}$ is the fitted continuum. If $I'_{0,i} \ne I_{0,i}$, then it may occur 
that $C' \ne C$.

The observed and laboratory wavelengths $C_{\alpha}$ and $C_{lab}$ of an atomic transition are related by
\begin{equation}
    \frac{1}{C_{\alpha}} = \frac{1}{C_{lab}} + q\left( \frac{\alpha^2}{\alpha_{lab}^2} -1 \right) \label{eq:da}
\end{equation}
where $\alpha$ is the fine structure constant in the measured gas cloud, $\alpha_{lab}$ is the terrestrial value, and $q$ is a transition-dependant sensitivity coefficient \cite{Dzuba1999b, Webb1999}. Eqs.\,\ref{eq:obscentroid} to \ref{eq:da} illustrate that a continuum placement uncertainty resulting in $C' \ne C$ causes a measurement in which we would measure $C'_{\alpha} \ne C_{\alpha}$, emulating a change in the fine structure constant. This effect has been ignored in many previous measurements of the fine structure constant in astronomical targets, yet should not be. We explore this in more detail in Section \ref{sec:method}.

In Section \ref{sec:PHL957}, the continuum modelling procedure is applied to a quasar spectrum obtained using the Ultra Violet Echelle Spectrograph (UVES) on the Very Large Telescope (VLT). By varying the continuum fitting knot spacings, i.e. generating different continuum models, we explore how variations in the adopted continuum model impact on measurements of the fine structure constant $\daa=(\alpha_z - \alpha_0)/\alpha_0$, where the subscripts $z, 0$ indicate redshift and the terrestrial value, and where the fine structure constant $\alpha = e^2/4\pi\epsilon_0\hbar c$ in SI units (Section \ref{sec:qimpact}). 

The analysis method for quasar absorption systems is necessarily very different to the white dwarf context. White dwarf photospheres produce a large number of narrow, generally weak, often single absorption lines, with known laboratory wavelengths. Whilst the high number density of lines (from multiple species) can result in some lines being blended, the complex kinematic structures seen in quasar absorption systems do not arise in white dwarf photospheres. In the quasar case, the profile modelling process returns a kinematic structure that depends on the continuum estimate provided. For that reason, it is important to derive independent models for each input continuum estimate when modelling quasar absorption systems.

In Section \ref{sec:PHL957} we present an {\sc ai-vpfit} measurement of $\daa$ in the $z_{abs}=1.7975$ absorption system towards the quasar PHL957. The fully AI method leaves $\alpha_z$ as a free parameter throughout the model building process, mandatory for an unbiased $\daa$ measurement \cite{Webb2022, Lee2023}. This new PHL957 $\daa$ measurement also, for the first time, includes additional free parameters to allow for a potential wavelength calibration linear distortion \cite{WebbVPFIT2021}, and checks (using multiple {\sc ai-vpfit} fits) for any possible model non-uniqueness effects \cite{Lee2020AI-VPFIT}.

\section{Continuum fitting procedure}\label{sec:method}

The method we use for obtaining an initial continuum has been presented in detail in \cite{Lee2024}. The advantage of (and reason for producing) that method is that it is automated, i.e. there is no interactive involvement beyond initially setting certain internal modelling parameters (or accepting default hard-coded values), and it is thus objective and reproducible. Since a comprehensive description exists elsewhere, only a brief summary is given here.

The method is based on cubic splines. Absorption (or emission) features in the spectrum are identified automatically and their associated pixels removed from the continuum modelling process. The procedure comprises 6 stages. In stage 1, the spectrum is  re-binned onto a finer grid  and smoothed using convolution with a Gaussian. In stage 2, a preliminary identification of pixels containing absorption (or emission) features is carried out (refined subsequently). The whole process is iterative but neither stage 1 or 2 are involved in further iterations. Since (following stages 1 and 2) the remaining data have gaps (where features were removed), knot positions are redistributed appropriately (in stage 3), to avoid over-fitting in spectral regions where only a few pixels remain. Stage 4 applies Gauss-Newton optimisation to iteratively solve for intensities at each knot position. Stages 5 and 6 involve further parameter refinements, tweaking the parameter updates and number of knots to miminise the objective function ($\chi^2$) optimally, and a final (but detailed) further refinement is carried out, testing for pixels that may have been incorrectly flagged as features, replacing them if so. The algorithm's parameters used in this analysis are as follows: smoothing FWHM $3\bar{x}$, $\zeta=3$, $k_{merge}=1/3$, $n=10$ (see \cite{Lee2024} for details).

\section{Astronomical and atomic data} \label{sec:qdata}

For the $\daa$ study in this paper, we use the $z_{abs} = 1.7975$ absorption system towards the well-studied $z_{em}=2.7$ quasar PHL957 (or Q0100+1300 or J010311+131617). This quasar has a long and distinguished history, with many papers devoted to it. Milestones include: its discovery (in a stellar survey) \cite{Haro1962}, M. Schmidt's identification as a quasar, reported in the spectroscopic study by \cite{Lowrance1972}, and then a detailed spectroscopic analysis in \cite{Coleman1976}. This absorption system was used for a $\daa$ measurement by \cite{Webb2011, King2012}. It was selected for the present study because we had previously found that it yields a small uncertainty and because of the availability of an existing high quality UVES spectrum: the VLT/UVES data for PHL957 used in this study were obtained by \cite{Zafar2013}. That paper gives the wavelength coverage, signal to noise, spectral resolution, and other details. The data extraction we use in this paper is from the compilation/re-reductions of \cite{Murphy2019}, in which continua are fitted to individual orders during data extraction, normalised and propagated through the co-addition procedure to form a final one-dimensional spectrum. The final one-dimensional spectrum provided in that compilation is thus normalised to unit continuum. We refer to that unit continuum as the ``original'' continuum. The compilation in \cite{Murphy2019} is large, comprising 467 reduced quasar spectra, and the provision of a basic continuum for each is extremely useful. However, those continua are relatively low order and not generally suitable for detailed studies involving line profile analysis, as we show in the present study.

In the context of the $\daa$ measurements made here, the most important atomic data parameters are the laboratory rest wavelengths $\lambda$ of the absorption transitions used, the oscillator strengths $f_{ik}$, damping constants $\Gamma$, sensitivity coefficients $q$, where the terminology is as given in the {\sc vpfit} user guide \cite{web:VPFIT, ascl:VPFIT2014}. The values of all atomic parameters used in the present study are as provided with {\sc vpfit} version 12.4 \citep{web:VPFIT}.

\section{Continuum models} \label{sec:qcontfits}

\begin{figure*}
\centering
\includegraphics[width=0.3\linewidth]{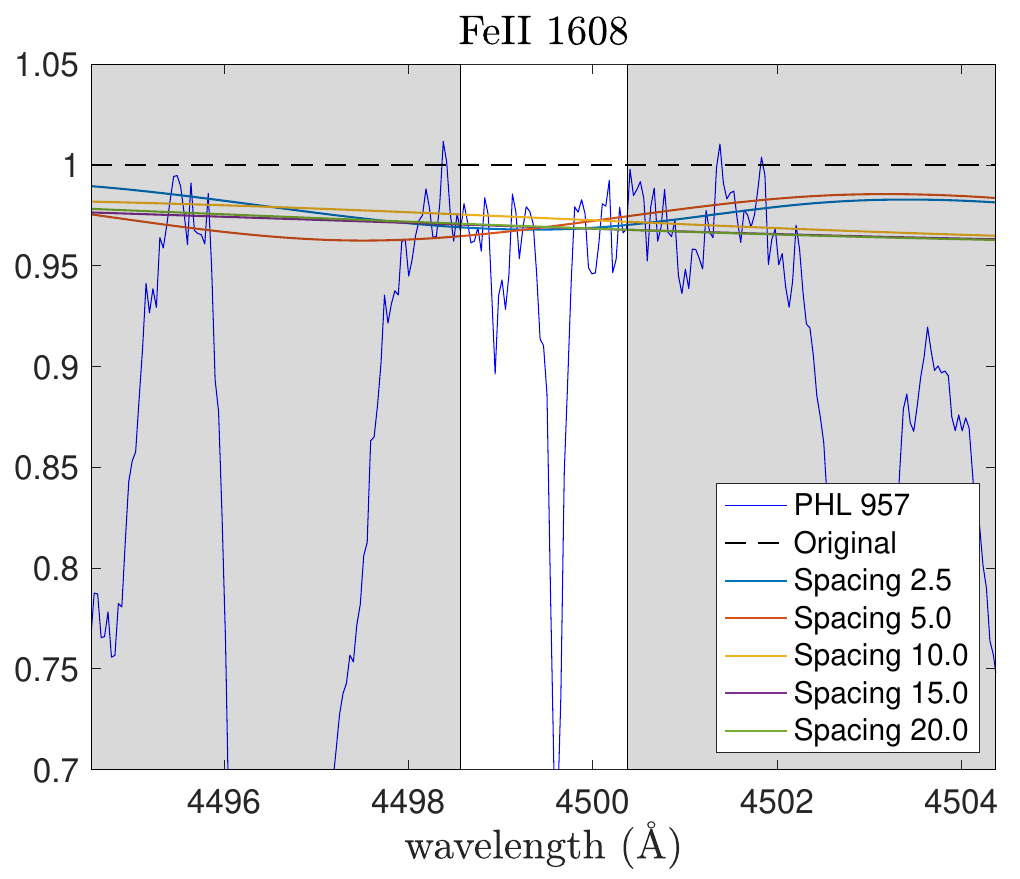}
\includegraphics[width=0.3\linewidth]{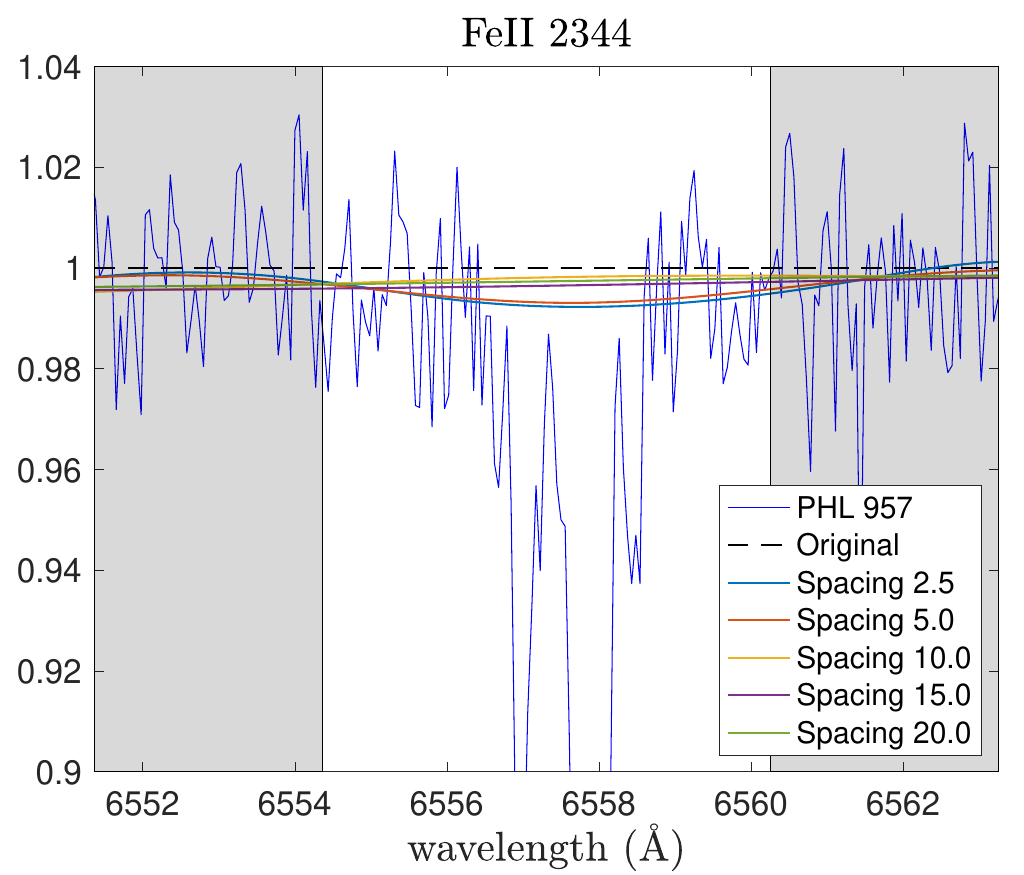}
\includegraphics[width=0.3\linewidth]{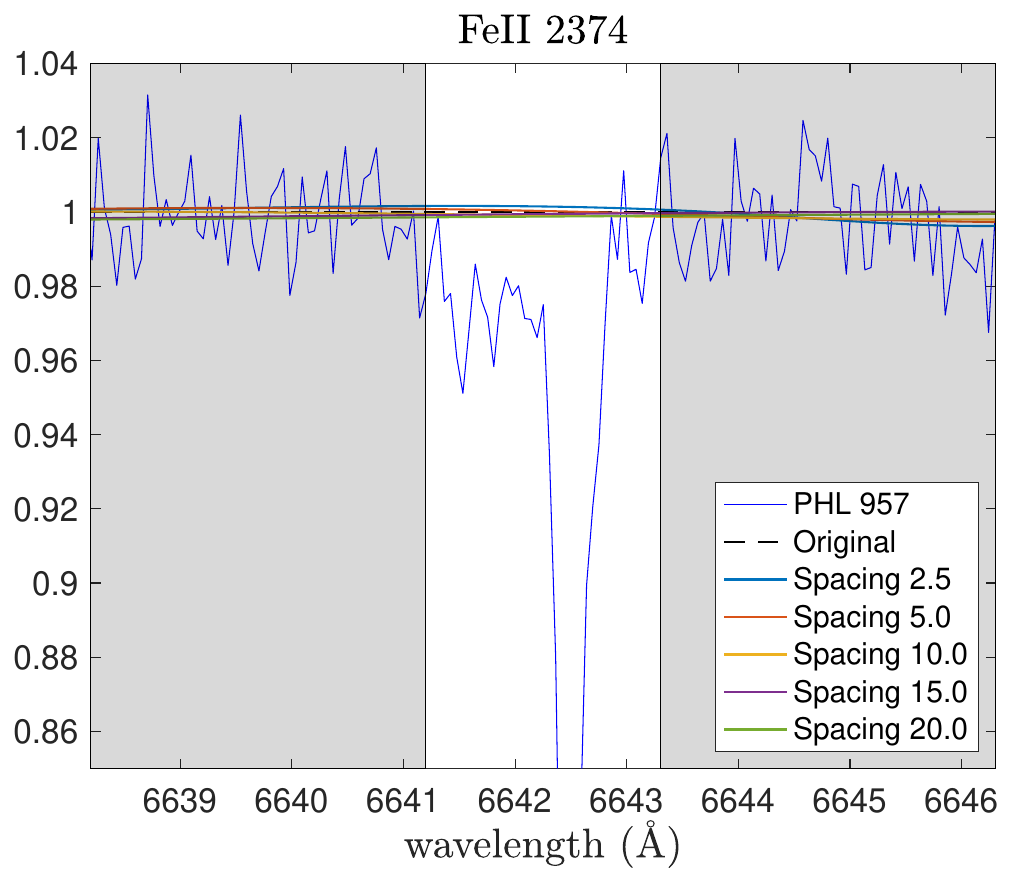}
\includegraphics[width=0.3\linewidth]{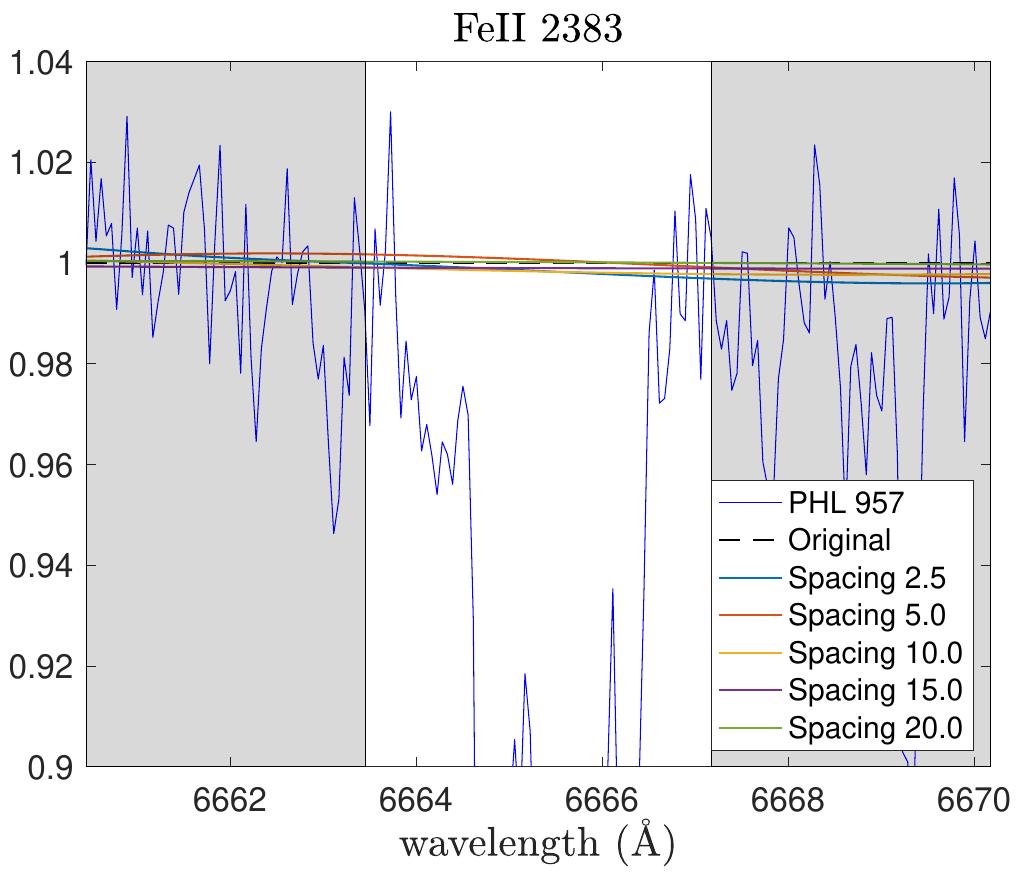}
\includegraphics[width=0.3\linewidth]{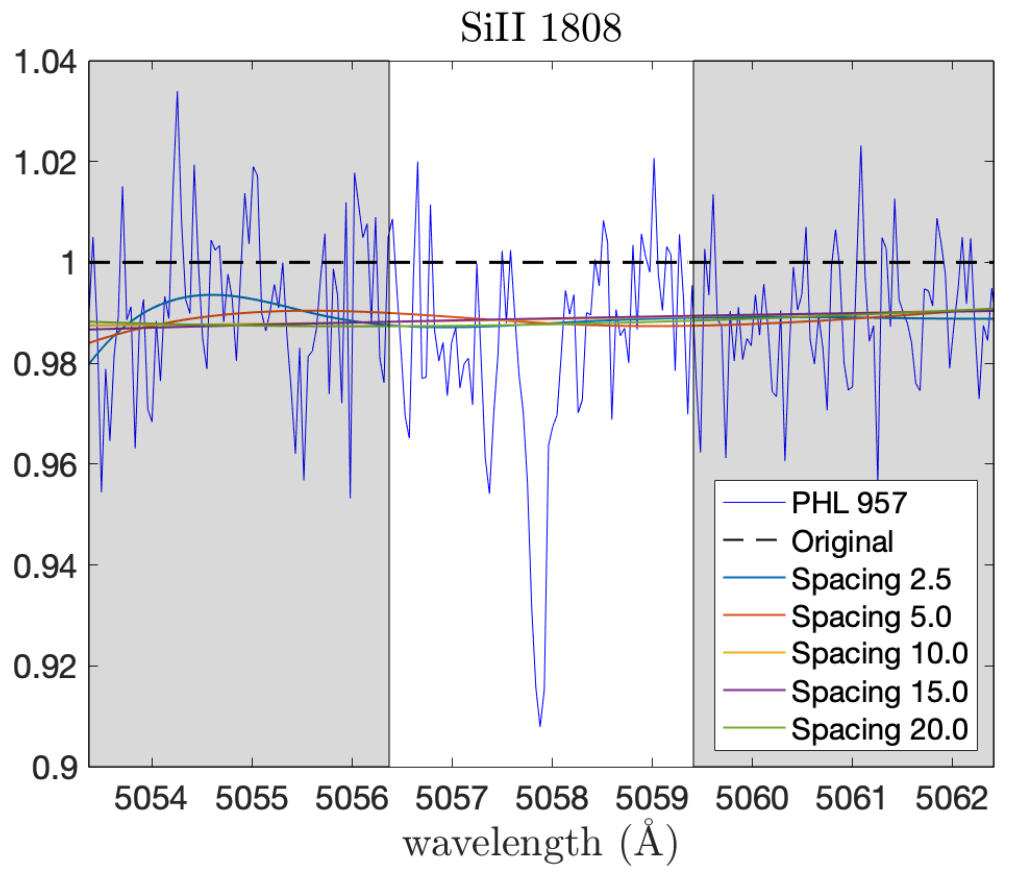}
\includegraphics[width=0.3\linewidth]{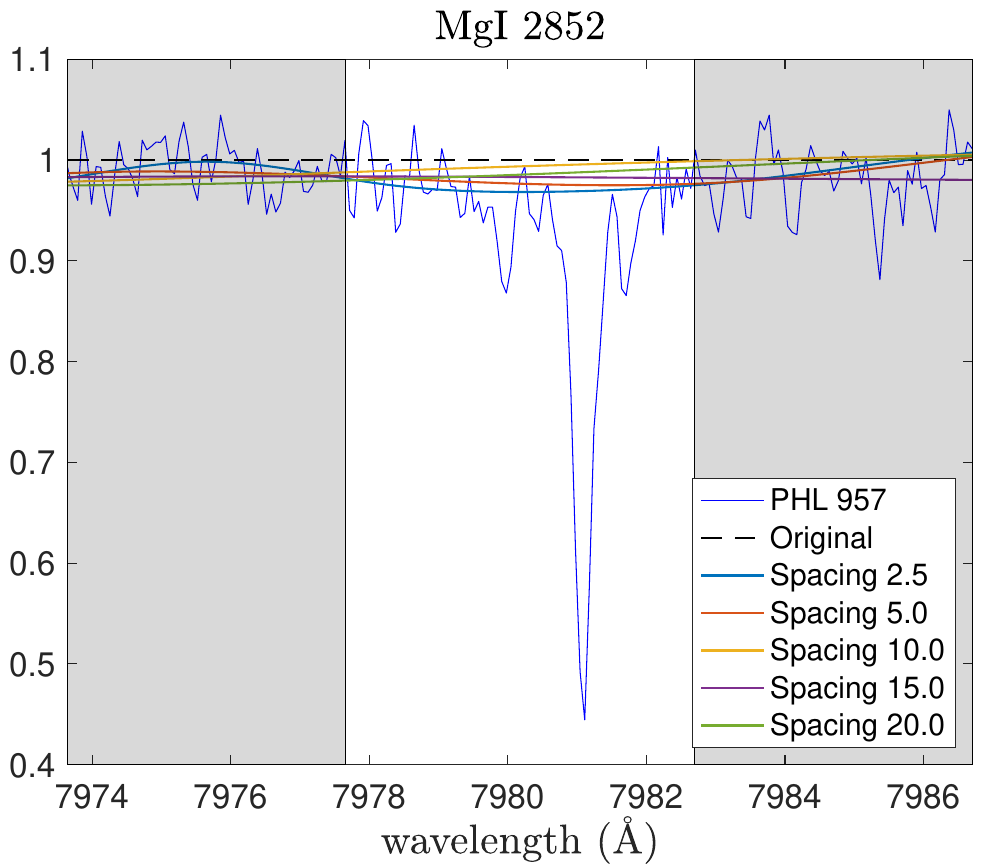}
\caption{PHL957 $z_{abs} = 1.7975$: continua (prior to refinement) for each of the six spectral fitting regions (5 new, with different knot spacings, and the original continuum derived during the data reduction procedures).
\label{fig:6transitions}}
\end{figure*}

For the $z_{abs}=1.7975$ absorption system towards PHL957 analysed here, the continuum level against which $\daa$ is measured is obtained as follows: \\
(i) first, prior to absorption line modelling, we derive a new {\it preliminary} continuum model, then \\
(ii) during absorption system modelling, we refine the local continuum estimates (independently for each spectral segment fitted), modifying the preliminary continuum $I_{0,\lambda}$ using a 2-parameter linear correction, so the final continuum becomes
\begin{equation}
I^{\prime}_{0,\lambda} = I_{0,\lambda} \left(c_l + c_s \left(\frac{\lambda}{\lambda_{ref}}-1 \right) \right), \label{eq:tilt}
\end{equation}
where $c_l$ and $c_s$ are constants specific to each of the six spectral regions fitted (whose wavelength ranges in this analysis are listed in the first column of Table \ref{tab:contparams}), $\lambda_{ref}$ is a reference wavelength within the fitting region in question, adopted as the central wavelength automatically in {\sc ai-vpfit}. The parameters $c_l$ and $c_s$ are solved for simultaneously with the other absorption system model parameters. Further details are given in the {\sc vpfit} user guide \cite{web:VPFIT} and also in \cite{WebbVPFIT2021}. Figure \ref{fig:6transitions} illustrates the six different continuum models for each spectral segment in the combined dataset.

\section{120 AI-VPFIT models of the $z_{abs}=1.7975$ system towards PHL957} \label{sec:PHL957}

The extensive calculations reported here are based on {\sc ai-vpfit} \citep{Lee2020AI-VPFIT}, which incorporates the {\sc vpfit} code, v12.4. Throughout the AI model construction process, the Spectroscopic Information Criterion (SpIC) is used for model selection \cite{Webb2021}.

\begin{table*}
\small
\renewcommand*{\arraystretch}{1.4}
\centering
\begin{tabular}{|l|c|c|c|c|c|c|}
\hline & \multicolumn{6}{c|}{\textbf{Knot spacing (\AA)}} \\ 
\hline
& \textbf{2.5}  & \textbf{5.0} & \textbf{10.0} & \textbf{15.0} & \textbf{20.0} & \textbf{Orig.} \\
\hline
\multicolumn{1}{|l|}{\textbf{Si\,{\sc ii}\,1526}} & 
$ 0.004 ( 1.3 \sigma)$ & $ 0.004 ( 1.3 \sigma)$ & $ 0.004 ( 1.2 \sigma)$ & $ 0.004 ( 1.1 \sigma)$ & $ 0.005 ( 1.4 \sigma)$ & $-0.007 (-2.1 \sigma)$ \\
\multicolumn{1}{|l|}{4498.56-4500.37} & 
$ 6.366 ( 0.4 \sigma)$ & $ 14.11 ( 1.0 \sigma)$ & $ 7.710 ( 0.5 \sigma)$ & $ 6.696 ( 0.4 \sigma)$ & $ 7.489 ( 0.5 \sigma)$ & $ 6.421  ( 0.4 \sigma)$ \\
\hline
\multicolumn{1}{|l|}{\textbf{Fe\,{\sc ii}\,1608}}  & 
$ 0.007 ( 3.2 \sigma)$ & 0.006 ( $ 2.9 \sigma)$ & $ 0.002 ( 1.0 \sigma)$ & $ 0.007 ( 3.1 \sigma)$ & $ 0.007 ( 3.0 \sigma)$ & $-0.012 (-2.5 \sigma)$ \\
\multicolumn{1}{|l|}{5056.37-5059.41} & 
$ 3.890 ( 0.2 \sigma)$ & $-18.67 (-1.0 \sigma)$ & $ 16.38 ( 0.9 \sigma)$ & $ 13.10 ( 0.7 \sigma)$ & $ 13.83 ( 0.8 \sigma)$ & $-7.230  (-0.2 \sigma)$ \\
\hline
\multicolumn{1}{|l|}{\textbf{Fe\,{\sc ii}\,2344}}  & 
$ 0.001 ( 0.4 \sigma)$ & $ 0.000 ( 0.0 \sigma)$ & $-0.003 (-2.1 \sigma)$ & $-0.002 (-1.1 \sigma)$ & $-0.002 (-1.3 \sigma)$ & $-0.004 (-2.0 \sigma)$ \\
\multicolumn{1}{|l|}{6554.36-6560.25} & 
$ 3.081 ( 0.5 \sigma)$ & $ 2.196 ( 0.4 \sigma)$ & $-0.427 (-0.1 \sigma)$ & $-0.468 (-0.1 \sigma)$ & $-1.551 (-0.2 \sigma)$ & $-2.313 (-0.4 \sigma)$ \\
\hline
\multicolumn{1}{|l|}{\textbf{Fe\,{\sc ii}\,2374}}  & 
$-0.010 (-3.9 \sigma)$ & $-0.008 (-3.5 \sigma)$ & $-0.007 (-3.1 \sigma)$ & $-0.008 (-3.2 \sigma)$ & $-0.007 (-2.7 \sigma)$ & $-0.008 (-3.3 \sigma)$ \\
\multicolumn{1}{|l|}{6641.20-6643.30} & 
$ 33.11 ( 1.3 \sigma)$ & $ 34.28 ( 1.3 \sigma)$ & $ 31.41 ( 1.2 \sigma)$ & $ 27.04 ( 1.1 \sigma)$ & $ 27.22 ( 1.1 \sigma)$ & $ 28.91 (1.2 \sigma)$ \\
\hline
\multicolumn{1}{|l|}{\textbf{Fe\,{\sc ii}\,2383}}  & 
$-0.006 (-1.6 \sigma)$ & $-0.008 (-2.5 \sigma)$ & $-0.006 (-1.8 \sigma)$ & $-0.007 (-2.1 \sigma)$ & $-0.007 (-1.6 \sigma)$ & $-0.005 (-1.0 \sigma)$ \\
\multicolumn{1}{|l|}{6663.45-6667.17} & 
$ 8.802 ( 0.5 \sigma)$ & $ 10.42 ( 0.7 \sigma)$ & $ 7.766 ( 0.5 \sigma)$ & $ 6.885 ( 0.5 \sigma)$ & $ 1.985 ( 0.1 \sigma)$ & $-7.139 (-0.3 \sigma)$ \\
\hline
\multicolumn{1}{|l|}{\textbf{Mg\,{\sc i}\,2851}}   & 
$ 0.007 ( 1.3 \sigma)$ & $ 0.001 ( 0.2 \sigma)$ & $-0.013 (-2.8 \sigma)$ & $-0.003 (-0.5 \sigma)$ & $-0.004 (-0.6 \sigma)$ & $-0.011 (-0.9 \sigma)$ \\
\multicolumn{1}{|l|}{7977.65-7982.70} & 
$-3.707 (-0.2 \sigma)$ & $ 7.379 ( 0.4 \sigma)$ & $-21.39 (-1.1 \sigma)$ & $-1.932 (-0.1 \sigma)$ & $-30.55 (-1.1 \sigma)$ & $-38.21 (-0.9 \sigma)$ \\
\hline
\end{tabular}    
\vspace{1em}\caption{{\sc ai-vpfit} continuum refinement parameters for each spectral segment used to fit $\daa$ in the PHL957 $z_{abs}=1.7975$ absorption system, for each initial continuum model. The leftmost column shows the six atomic species used, their rest-frame wavelengths, and underneath, the observed-frame fitting ranges in {\AA}. The header shows knot spacings in {\AA}. The rightmost column, ``Orig.'', uses the continuum provided with the extracted spectrum (see Section \ref{sec:qdata}). Each box shows $\langle (c_l-1) \rangle$ (upper row) and $\langle c_s \rangle$ (lower). The quantities in brackets illustrate the statistical significance (and signs of) of the deviations from $c_l =1$ and $c_s =0$, i.e. $\langle (c_l-1) /\sigma(c_l) \rangle$ and $\langle c_s/\sigma(c_s) \rangle$. \label{tab:contparams}}
\end{table*}

Many previous $\daa$ measurements in the literature have not applied step (ii) above, i.e. the assumption has often been made that the original continuum is good so that no additional corrections were necessary, an assumption we show to be incorrect in this paper. Many published $\daa$ measurements thus have quoted final error budgets that are too small; high precision $\daa$ spectroscopic measurements should {\it always} account for this source of uncertainty. \\

\subsection{Modelling wavelength distortion} \label{sec:distortion}

The focus of the present paper concerns the impact of continuum placement uncertainty on $\daa$. However, previous detailed studies of VLT/UVES and Keck/HIRES spectra have revealed an approximately linear wavelength distortion effect. This effect should not be ignored here since we aim to derive a realistic overall $\daa$ uncertainty. 

An explicit search for possible wavelength distortions in high-resolution quasar spectra was first reported in \cite{Molaro2008}, who correlated the reflected solar spectrum (obtained from asteroid spectra observed using UVES on the VLT) with independent solar calibrations. The result of that initial study found no evidence for long-range wavelength distortion for VLT/UVES spectra. However, using higher precision data, \cite{Rahmani2013} showed that in fact long-range wavelength distortions can arise in UVES spectra, and that, {\it for constant observational instrument settings}, the distortion pattern can be approximated using a simple linear relationship between velocity shift and observed wavelength. However, a single linear distortion correction does not generally apply to quasar absorption system $\daa$ measurements (as was done in \cite{Whitmore2015}) because, in many cases, final spectra are formed by co-adding multiple exposures with different observational settings. Instead, one must derive an appropriately weighted combination of shifted linear distortion patterns, taking into account instrument settings for all contributing exposures \cite{Dumont2017}. Additional free parameters for modelling potential distortion have been incorporated into {\sc ai-vpfit} and {\sc vpfit} version 12.4 \cite{web:VPFIT}. This problem is also discussed in Appendix B of \cite{WebbVPFIT2021}.

The distortion slopes measured for PHL957 in our {\sc ai-vpfit} calculations are given in Table \ref{tab:distortion_parm}, which show no significant evidence for long-range wavelength distortion. The first search for the presence of long-range wavelength distortion in the PHL957 spectrum used in the present paper was carried out as part of the PhD thesis research by \cite{Dumont2018}. In that study, the analysis was simpler; the two $z_{abs}=1.7975$ absorption system models given in \cite{King2012} were assumed, one based on thermal line broadening, the other based on turbulent broadening, and distortion parameters were solved for external to {\sc vpfit}. The \cite{Dumont2017} approach is therefore slightly different to the {\sc ai-vpfit} analyses described in this paper, because here we use more realistic compound line broadening and solve for velocity structure and distortion parameters simultaneously. Nevertheless, interestingly, the slope parameters given in \cite{Dumont2017} for the $z_{abs}=1.7975$ absorption system towards PHL957 are $\gamma = 0.09 \pm 0.07$ (thermal broadening), $0.26 \pm 0.08$ (turbulent broadening), and $0.26 \pm 0.07$ (method of moments) m\,s$^{-1}$\,\AA$^{-1}$, consistent with the set of mean {\sc ai-vpfit} results given in Table \ref{tab:distortion_parm}, which lie in the range $0.14\pm10 < \gamma < 0.18\pm10$ m\,s$^{-1}$\,\AA$^{-1}$. Since the distortion parameters are included in the {\sc ai-vpfit} covariance matrix, distortion uncertainty propagates correctly to the final $\daa$ uncertainty.

\begin{table*}
\renewcommand*{\arraystretch}{1.8}
\centering
\begin{tabular}{|c|c|c|c|c|c|c|}
\hline
\textbf{Knots (\AA)} & \textbf{2.5} & \textbf{5.0} & \textbf{10.0} & \textbf{15.0} &\textbf{20.0} & \textbf{Orig.} \\
\hline
 1  & $0.213 \pm 0.092$ & $0.160 \pm 0.088$ & $0.160 \pm 0.089$ & $0.119 \pm 0.107$ & $0.154 \pm 0.087$ & $0.062 \pm 0.106$ \\ \hline 
 2  & $0.111 \pm 0.108$ & $0.221 \pm 0.091$ & $0.112 \pm 0.107$ & $0.210 \pm 0.089$ & $0.126 \pm 0.106$ & $0.182 \pm 0.107$ \\ \hline 
 3  & $0.085 \pm 0.109$ & $0.241 \pm 0.089$ & $0.262 \pm 0.088$ & $0.262 \pm 0.130$ & $0.100 \pm 0.106$ & $0.150 \pm 0.126$ \\ \hline 
 4  & $0.249 \pm 0.089$ & $0.092 \pm 0.104$ & $0.084 \pm 0.109$ & $0.262 \pm 0.125$ & $0.235 \pm 0.089$ & $0.129 \pm 0.099$ \\ \hline 
 5  & $0.181 \pm 0.092$ & $0.090 \pm 0.109$ & $0.081 \pm 0.110$ & $0.093 \pm 0.108$ & $0.085 \pm 0.109$ & $0.179 \pm 0.086$ \\ \hline 
 6  & $0.092 \pm 0.102$ & $0.244 \pm 0.089$ & $0.183 \pm 0.131$ & $0.233 \pm 0.088$ & $0.251 \pm 0.089$ & $0.191 \pm 0.126$ \\ \hline 
 7  & $0.238 \pm 0.089$ & $0.101 \pm 0.108$ & $0.109 \pm 0.108$ & $0.086 \pm 0.105$ & $0.068 \pm 0.107$ & $0.172 \pm 0.090$ \\ \hline 
 8  & $0.176 \pm 0.096$ & $0.340 \pm 0.112$ & $0.083 \pm 0.109$ & $0.080 \pm 0.106$ & $0.303 \pm 0.106$ & $0.134 \pm 0.102$ \\ \hline 
 9  & $0.249 \pm 0.089$ & $0.074 \pm 0.106$ & $0.214 \pm 0.091$ & $0.148 \pm 0.086$ & $0.232 \pm 0.090$ & $0.114 \pm 0.102$ \\ \hline 
10  & $0.168 \pm 0.094$ & $0.320 \pm 0.104$ & $0.215 \pm 0.092$ & $0.085 \pm 0.111$ & $0.096 \pm 0.098$ & $0.162 \pm 0.084$ \\ \hline 
11  & $0.079 \pm 0.110$ & $0.078 \pm 0.110$ & $0.107 \pm 0.104$ & $0.094 \pm 0.110$ & $0.068 \pm 0.108$ & $0.184 \pm 0.090$ \\ \hline 
12  & $0.089 \pm 0.112$ & $0.080 \pm 0.110$ & $0.073 \pm 0.106$ & $0.231 \pm 0.090$ & $0.257 \pm 0.088$ & $0.273 \pm 0.128$ \\ \hline 
13  & $0.175 \pm 0.091$ & $0.098 \pm 0.102$ & $0.135 \pm 0.097$ & $0.163 \pm 0.090$ & $0.193 \pm 0.090$ & $0.344 \pm 0.106$ \\ \hline 
14  & $0.253 \pm 0.089$ & $0.235 \pm 0.090$ & $0.257 \pm 0.088$ & $0.241 \pm 0.089$ & $0.247 \pm 0.091$ & $0.299 \pm 0.110$ \\ \hline 
15  & $0.216 \pm 0.092$ & $0.117 \pm 0.111$ & $0.098 \pm 0.107$ & $0.191 \pm 0.092$ & $0.087 \pm 0.110$ & $0.210 \pm 0.089$ \\ \hline 
16  & $0.216 \pm 0.091$ & $0.107 \pm 0.097$ & $0.110 \pm 0.109$ & $0.227 \pm 0.122$ & $0.085 \pm 0.110$ & $0.078 \pm 0.106$ \\ \hline 
17  & $0.210 \pm 0.092$ & $0.217 \pm 0.088$ & $0.201 \pm 0.122$ & $0.067 \pm 0.110$ & $0.119 \pm 0.110$ & $0.176 \pm 0.095$ \\ \hline 
18  & $0.231 \pm 0.087$ & $0.089 \pm 0.105$ & $0.062 \pm 0.105$ & $0.190 \pm 0.090$ & $0.086 \pm 0.118$ & $0.256 \pm 0.156$ \\ \hline 
19  & $0.120 \pm 0.107$ & $0.241 \pm 0.089$ & $0.158 \pm 0.092$ & $0.194 \pm 0.090$ & $0.196 \pm 0.097$ & $0.180 \pm 0.093$ \\ \hline 
20  & $0.054 \pm 0.111$ & $0.108 \pm 0.106$ & $0.073 \pm 0.109$ & $0.068 \pm 0.107$ & $0.075 \pm 0.109$ & $0.167 \pm 0.095$ \\ \hline 
Means& $0.170 \pm 0.097$ & $0.163 \pm 0.101$ & $0.139 \pm 0.104$ & $0.162 \pm 0.102$ & $0.153 \pm 0.101$ & $0.182 \pm 0.105$ \\
\hline
\end{tabular}    
\vspace{1em}\caption{Best fit distortion slopes for the $z_{abs}=1.7975$ absorption system towards PHL957, in m\,s$^{-1}$\,\AA$^{-1}$, for all 120 {\sc ai-vpfit} absorption system models. See Section \ref{sec:distortion} for details.
\label{tab:distortion_parm}}
\end{table*}

\subsection{Impact of continuum uncertainty on \texorpdfstring{$\daa$}{daa} measurement with {\sc vpfit} continuum refinement} \label{sec:qimpact}

Table \ref{tab:contparams} gives the continuum refinement parameters and their associated 1$\sigma$ uncertainties, for all six spectral segments used. Each box in the table contains $(c_l - 1)/\sigma(c_l)$ (upper row) and $c_s/\sigma(c_s)$ (lower), i.e. the quantities provide deviations from no-correction as multiples of 1 standard deviation, and thus illustrate whether or not the non-linear least squares procedure deemed the parameter to be statistically necessary. The uncertainties $\sigma(c_l)$ and $\sigma(c_s)$ are obtained from {\sc vpfit's} parameter covariance matrix. Table \ref{tab:contparams} shows that 8 out of 36 continuum corrections require corrections with a significance level of $3\sigma$ or larger, associated with 2 of the 6 spectral fitting regions, suggesting that corrections this large might be commonplace. This is interesting because it illustrates that even if the original continuum appears good (visually), as is the case for PHL957, local adjustments (i.e. for each fitting region) are likely to be needed.

Table \ref{tab:contparams} also shows that {\sc ai-vpfit} required an increasing number of components as the knot spacing increases, the largest number being required when {\sc ai-vpfit} was supplied with the original continuum. This trend is unsurprising because as the knot spacing is increased, the spline model tends to the (coarser) original continuum. The variation in the number of components does not indicate a preference for any particular continuum, but it does provide independent evidence that there is a source of uncertainty associated with continuum modelling that must be taken into account in the overall $\daa$ error budget. However, the results illustrated in Table \ref{tab:contparams} do not show whether these local continuum adjustments have any impact on the measured $\daa$. We thus next examine the corresponding changes in $\daa$, with and without these corrections.

\begin{figure*}
\centering
\includegraphics[width=0.9\linewidth]{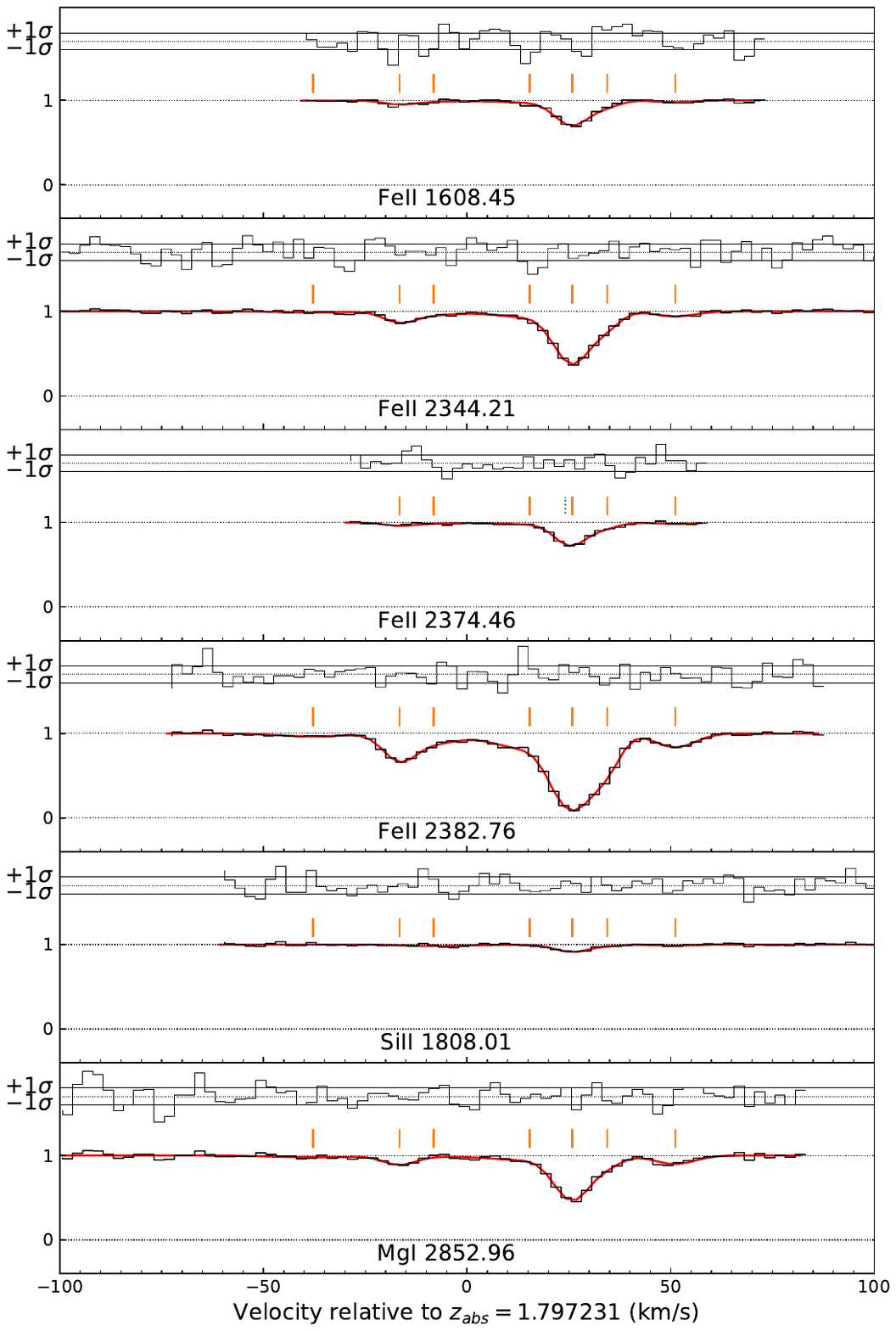}
\caption{Best AI-VPFIT/SpIC model with continuum knot spacing 10{\AA} \label{fig:spic_best}}
\end{figure*}

\begin{figure*}
\centering
\includegraphics[width=0.3\linewidth]{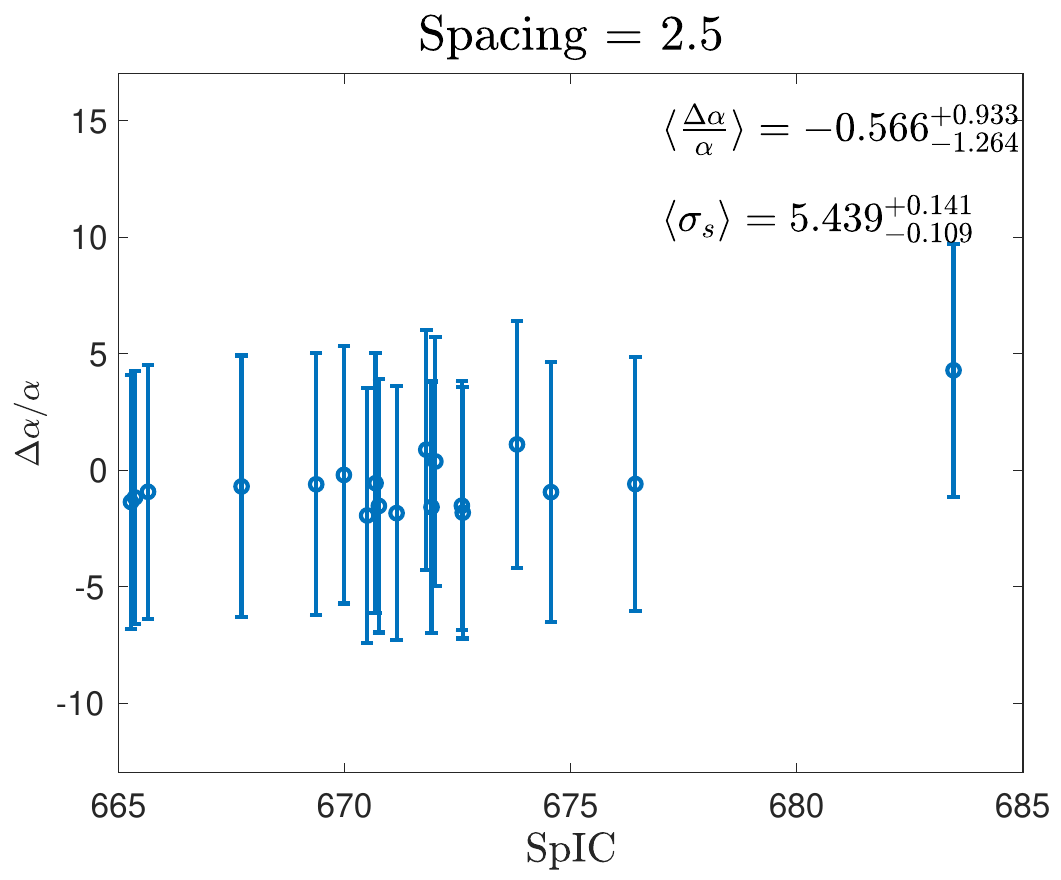}
\includegraphics[width=0.3\linewidth]{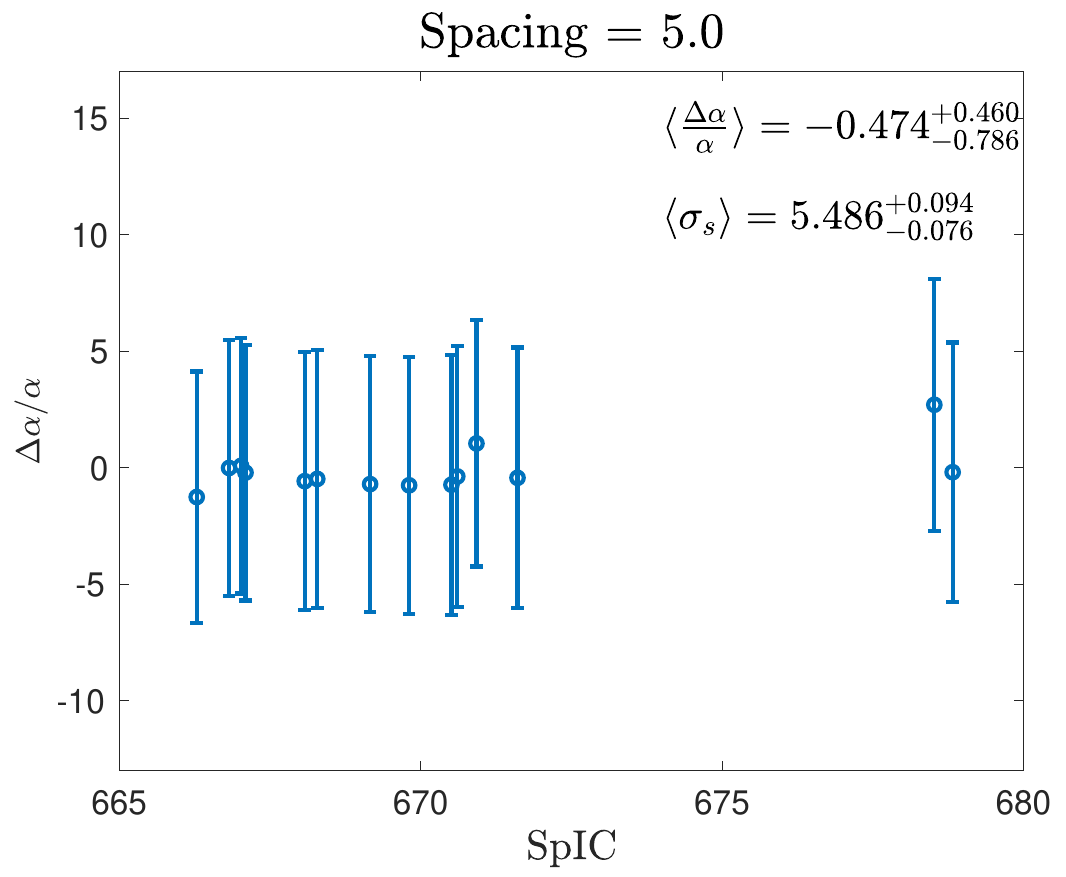}
\includegraphics[width=0.3\linewidth]{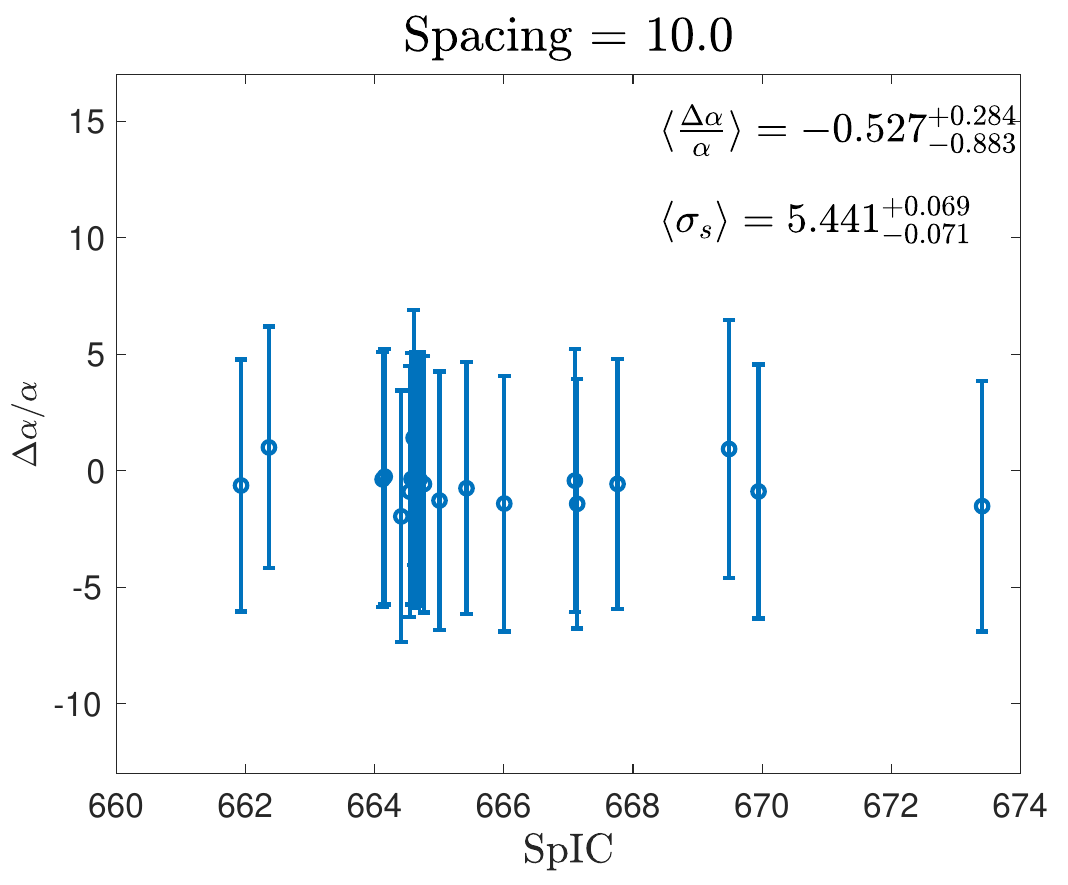}
\includegraphics[width=0.3\linewidth]{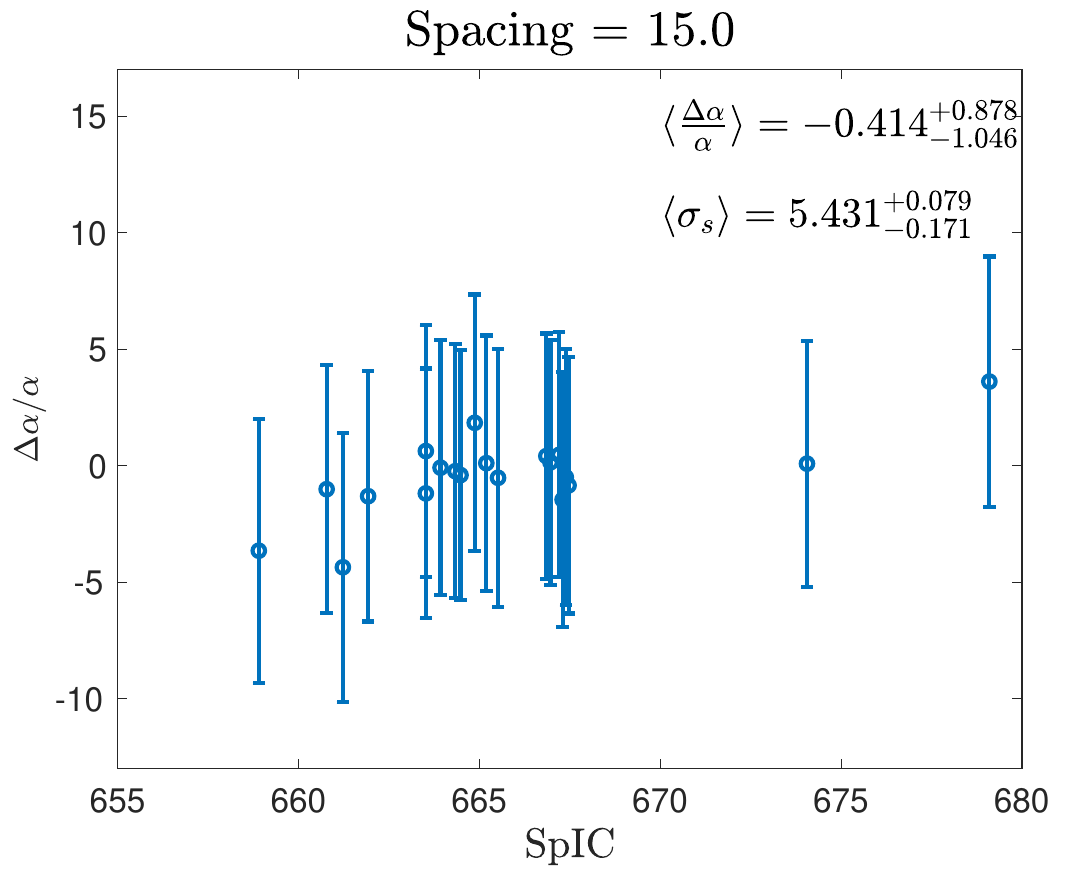}
\includegraphics[width=0.3\linewidth]{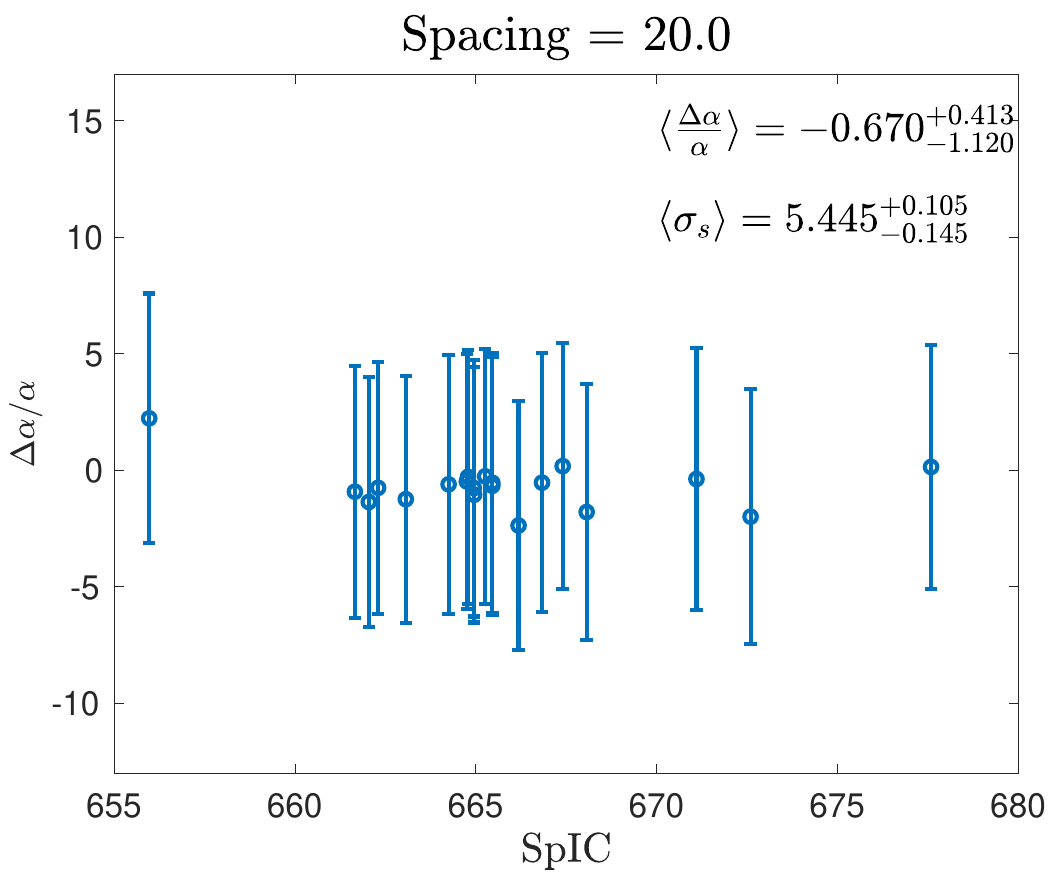}
\includegraphics[width=0.3\linewidth]{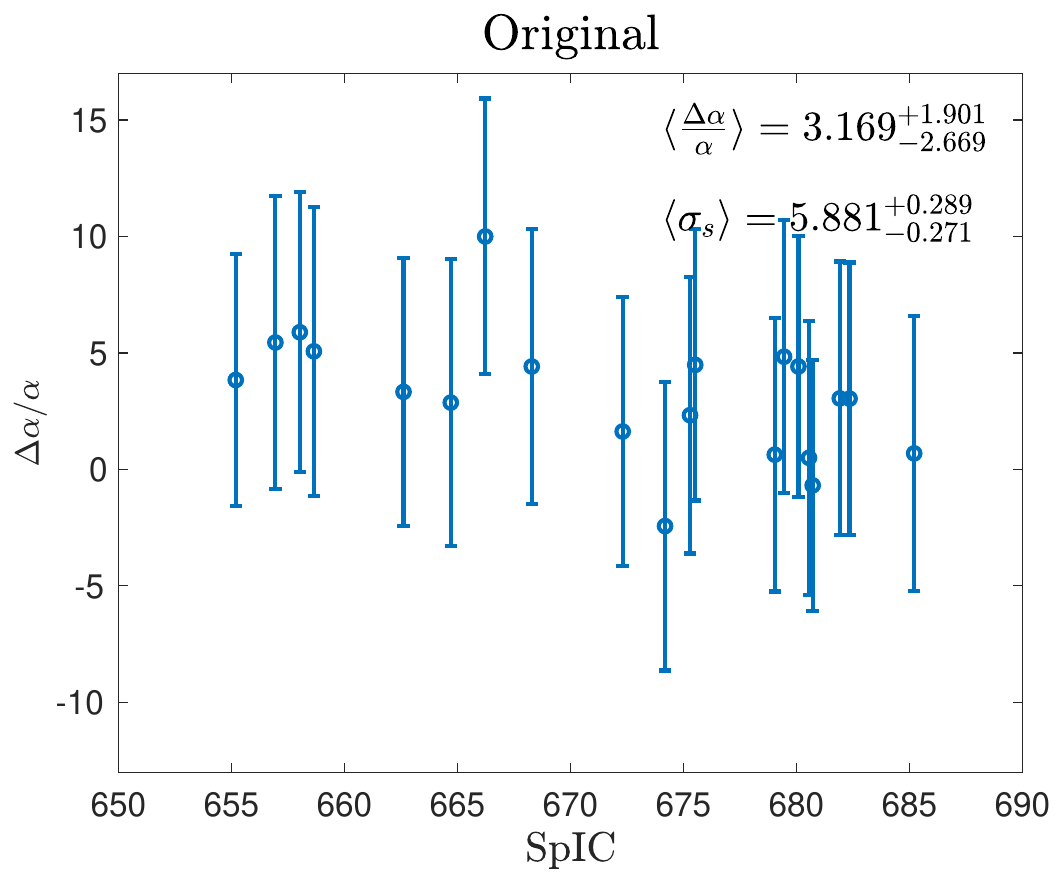}
\caption{120 $\daa$ measurements from {\sc ai-vpfit} models of the $z_{abs} = 1.7975$ absorption system towards PHL957 (20 independent {\sc ai-vpfit} $\daa$ measurements for each of 6 continua). Free parameters to allow for any possible linear wavelength distortion were also included. See Section \ref{sec:qimpact} for details. \label{fig:4qresults}}
\end{figure*}

The calculations comprise: 20 independent models for each of 6 continuum $\daa$ measurements (2.5, 5.0, 10.0, 15, 20 {\AA} knot spacings, plus the original continuum) i.e. we have computed a total of 120 {\sc ai-vpfit} models for the $z_{abs}=1.7975$ absorption system. Figure \ref{fig:spic_best} illustrates one example model (smallest SpIC, continuum knot spacing 10{\AA}). Figure \ref{fig:4qresults} illustrates the 120 individual $\daa$ measurements. For each model, continuum parameters were included, allowing the initial continuum model to refine. The plotted error bar illustrates the statistical (covariance matrix) error. Insets show $\langle\daa\rangle$, the mean over 20 measurements and its associated point to point scatter for each set of 20 measurements $\langle\sigma_s\rangle$. This additional $\daa$ scatter is caused by two effects: first, as the last row of Table \ref{tab:contparams} shows, small continuum variations give rise to slightly different kinematic structures with different numbers of absorption components. Second, it is well known that non-linear least squares methods with many free parameters generally suffer from convergence issues e.g. \cite{Webb2024}. We have not attempted to quantify the balance between these two effects. Table \ref{tab:qimpact_withrefine} summarises the results illustrated in Figure \ref{fig:4qresults}. The upper and lower limits are the 1$\sigma$ uncertainties about the illustrated measurement, formed from the quadrature addition of the symmetric statistical uncertainty and the asymmetric convergence scatter. 

\begin{table*}
\renewcommand*{\arraystretch}{1.8}
\centering
\begin{tabular}{|l|c|c|c|c|c|c|}
\hline
\textbf{Knots (\AA)} & \textbf{2.5} & \textbf{5.0} & \textbf{10.0} & \textbf{15.0} & \textbf{20.0} & \textbf{Orig.} \\
\hline
\textbf{$\langle\daa\rangle$} & $-0.57^{+5.5}_{-5.6}$ & $-0.47^{+5.5}_{-5.5}$ & $-0.53^{+5.4}_{-5.5}$ & $-0.41 ^{+5.5}_{-5.5}$ & $-0.67^{+5.5}_{-5.6}$ & $3.2^{+6.1}_{-6.4}$ \\
\hline
\textbf{$\langle$ N $\rangle$} & $7.75 \pm 1.07$ & $7.70 \pm 1.03$ & $8.05 \pm 0.76$ & $8.05 \pm 0.83$ & $8.00 \pm 1.34$ & $9.65 \pm 1.35$ \\
\hline
\end{tabular}    
\vspace{1em}\caption{PHL957 $z_{abs}=1.7975$ with continuum refinement: the impact of using different quasar continuum models on measurements of $\daa$ (in units of $10^{-6}$). Six different continua are used (5 knot spacings plus original continuum). The automated continuum fitting parameters are all described in \cite{Lee2024}. Six transitions are modelled simultaneously: Fe\,{\sc ii}\,1608, Fe\,{\sc ii}\,2344, Fe\,{\sc ii}\,2374, Fe\,{\sc ii}\,2383, Si\,{\sc ii}\,1808, and Mg\,{\sc i}\,2853 {\AA}. The asymmetric $\daa$ uncertainties are formed from the symmetric statistical uncertainty in quadrature addition with the smaller asymmetric non-uniqueness scatter. The lowest row shows the mean number of heavy element absorption components (with 1$\sigma$ error). Averages are taken over the 20 {\sc ai-vpfit} models for each knot setting in all cases (see Figure 3 insets). See Section \ref{sec:qimpact} for further details. \label{tab:qimpact_withrefine}}
\end{table*}

\begin{table*}
\renewcommand*{\arraystretch}{1.8}
\centering
\begin{tabular}{|l|c|c|c|c|c|c|}
\hline
\textbf{Knots (\AA)} & \textbf{2.5} & \textbf{5.0} & \textbf{10.0} & \textbf{15.0} &\textbf{20.0} & \textbf{Orig.} \\
\hline
$\langle\daa\rangle$ & $0.91^{+5.6}_{-6.3}$ & $1.2^{+5.7}_{-5.8}$ & $1.0^{+5.7}_{-5.7}$ & $-0.77^{+5.7}_{-5.7}$ & $-1.0^{+5.6}_{-5.6}$ & $4.9^{+6.4}_{-6.6}$ \\
\hline
\textbf{$\langle$ N $\rangle$} & $8.00 \pm 0.79$ & $8.75 \pm 0.79$ & $9.20 \pm 0.70$ & $7.55 \pm 0.51$ & $7.35 \pm 0.49$ & $11.25 \pm 1.21$ \\
\hline
\end{tabular}
\vspace{1em}\caption{PHL957 $z_{abs}=1.7975$, without continuum refinement: same as Table \ref{tab:qimpact_withrefine} but no free continuum model parameters were included in {\sc ai-vpfit} modelling i.e. only the initial continuum models were used. $\daa$ is in units of $10^{-6}$. The mean number of components is formed from an average over 20 models in each case.} \label{tab:qimpact_norefine}
\end{table*}

Table \ref{tab:qimpact_norefine} gives the analogous results where {\it no} continuum refinement parameters were included. Comparing this table with Table \ref{tab:qimpact_withrefine} therefore provides a check on whether, for this particular quasar measurement at least, continuum refinement had any effect. In all 6 comparisons, we see that $\daa$ does indeed change when continuum refinement is applied. The magnitude of the mean shift is $1.2 \times 10^{-6}$, approximately 1/5 of the statistical $\daa$ error bar, which itself correctly allows for the presence of the continuum refinement parameters since the latter is derived from the model covariance matrix.

\subsection{What causes the discrepant result obtained using the original continuum?} \label{sec:contreplace}

Visual inspection (black continuous line in Figure \ref{fig:6transitions}) suggests that the initial, ``unrefined'', continuum, ``Orig.'' in Tables \ref{tab:qimpact_withrefine} and \ref{tab:qimpact_norefine}, is reasonable. However, some deviations between that original continuum and the 5 new (unrefined) models are noticeable, particularly across two transitions, FeII 2344 and SiII 1808{\AA}. Nevertheless, the {\sc ai-vpfit} results illustrated in Figure \ref{fig:4qresults} and summarised in Table \ref{tab:qimpact_withrefine} show that, even after refinement (i.e. {\sc ai-vpfit} optimisation of the $c_s$ and $c_l$ parameters for each spectral region), the refined original continuum gives $\langle\daa\rangle = 3.2^{+6.1}_{-6.4} \times 10^{-6}$, whereas all other measurements (based on a new initial continuum model derived using the method described in Section \ref{sec:method}) are around $-0.5 \times 10^{-6}$, with very little scatter. Even after refinement of all 6 continua, i.e. including the original, the discrepancy between the original continuum and refined continuum results corresponds to $\sim$67\% of the best statistical (i.e. covariance matrix) uncertainty and is therefore important. 

To investigate the source of the effects described in the preceding paragraph, we carried out an additional set of 120 {\sc ai-vpfit} calculations, with continuum refinement, in which the initial continua for 5 (out of of 6) spectral segments were the original ones, whilst the continuum for one remaining spectral segment at a time was replaced with the locally refined continuum. This was done for one specific knot spacing only (5{\AA}, because calculation times are fairly long). The results from these 120 new models are given in Table \ref{tab:contreplace}, which show the strongest impact arises when the Fe\,{\sc ii} 1608{\AA} continuum is replaced, suggesting the difference between the original and refined continua plays an more important role in this wavelength region (see Fig.\ref{fig:6transitions}). Interestingly, Si\,{\sc ii} 1808{\AA} also displays a similarly large offset but this appears not to impact substantially in this case.

\begin{table*}
\renewcommand*{\arraystretch}{1.8}
\centering
\begin{tabular}{|c|c|c|c|c|c|c|}
\hline
\textbf{Replacement:} & Fe\,{\sc ii}\,1608 & Fe\,{\sc ii}\,2344 & Fe\,{\sc ii}\,2374 & Fe\,{\sc ii}\,2383 & Si\,{\sc ii}\,1808 & Mg\,{\sc i}\,2853 \\
\hline
$\langle\daa\rangle$: & $-0.49^{+5.5}_{-5.5}$ & $1.9^{+6.0}_{-6.1}$ & $2.8^{+6.4}_{-6.5}$ & $3.5^{+6.3}_{-6.5}$ & $3.0^{+6.2}_{-6.5}$ & $2.6^{+6.0}_{-6.1}$\\
\hline
\end{tabular}
\vspace{1em}\caption{PHL957 $z_{abs}=1.7975$: continuum replacement test carried out using the 5{\AA} knot spacing continuum model. The asymmetric $\daa$ uncertainties are formed from the symmetric statistical uncertainty in quadrature addition with the smaller asymmetric non-uniqueness scatter. Each $\langle\daa\rangle$ value is the mean over 20 independently formed models. $\daa$ is in units of $10^{-6}$. With no replacement, the refined original continuum gives $\langle\daa\rangle = 3.2^{+6.1}_{-6.4} \times 10^{-6}$. See Section \ref{sec:contreplace} for details.}
\label{tab:contreplace}
\end{table*}

\section{Conclusions and discussion}\label{sec:conclusions}

The calculations presented in this paper have shown the importance of careful continuum preparation, {\it and} that the continuum model needs to be refined further, simultaneous with absorption system model optimisation, else a significant additional source of uncertainty may be added to a $\daa$ measurement. This uncertainty has been ignored in many, even most, previous measurements. We have studied only a single absorption system, so do not know whether our results apply more broadly (although there is no reason to think they do not). The work presented in this paper should not be interpreted as a criticism of the original continua provided in the compilation of \cite{Murphy2019}; that initial basic continuum is useful and provides an invaluable starting point for the more refined calculations described in this paper. With the caveat above, we draw the following conclusions, in the context of minimising systematic errors associated with $\daa$ measurements:

\begin{enumerate}
\item Continuum uncertainty (as described by Eq.\,\eqref{eq:tilt}) significantly alters $\daa$ estimates. By ``significantly'', we mean here that the additional uncertainty on $\daa$ associated with continuum placement uncertainty is a non-negligible fraction of the statistical uncertainty on $\daa$. Thus, when attempting to continuum-fit high resolution quasar spectra, it is desirable to have an objective and reproducible method that avoids interactive human decision making as far as possible.
\item In this study, we take a two-stage approach, (i) prior to any absorption system modelling, fit new continua, and then (ii) refine those continua simultaneous with absorption system modelling. In most previous searches for varying fundamental constants, this has not been done. Our findings suggest that if this is done, any systematic on the inferred $\daa$ measurement may be reduced to a negligible level (i.e. far below the covariance matrix uncertainty) although further studies of other systems are needed to confirm this.
\item The 5 new models we derive (after fine-tuning, Table \ref{tab:qimpact_withrefine}) produce consistent $\daa$ results. We do not know if that consistency is generally true, but this one result clearly motivates checking for each new $\daa$ measurement made and emphasises the necessity for a carefully derived continuum before attempting to model absorption profiles. Perhaps counter-intuitively, the $\daa$ uncertainty estimate tends to decrease slightly, rather than increase, when additional continuum parameters are included in the modelling. The explanation may in part be that an improved local continuum often (but not always) tends to reduce the number of absorption line components in the best-fit model. 
\item In the case studied here, the original (unrefined) supplied continuum (despite appearing visually to be quite good) gives rise to a large systematic offset in the measured $\daa$, compared to the higher-order continuum models we calculated, with or without fine-tuning.
\item The $z_{abs}=1.7975$ absorption system towards PHL957 appears not to suffer from non-uniqueness problems \cite{Lee2021}, as Figure \ref{fig:4qresults} shows. This absorption system therefore appears to be a ``good'' system for such measurements in that specific sense. Our example or illustrative result is $\daa = -0.53^{+5.4}_{-5.5} \times 10^{-6}$. However, during the course of the study reported in this paper, a parallel analysis was done to explore the impact of isotopic abundance assumptions on the inferred $\daa$ measurement \cite{Webb2024mystery}. Magnesium is of particular concern in this respect because lines from its stable isotopes are more widely spaced in wavelength than most other species general used for $\daa$ measurements \citep{Berengut2005, Salumbides2006}. This means that modelling line positions with profiles formed using relative isotopic abundances that are incorrect can produce artificial line shifts, emulating $\daa \ne 0$ \citep{Webb1999}. The effect was subsequently investigated in several studies, e.g. \cite{Kozlov2004, Ashenfelter2004PRL, Fenner2005, Agafonova2011, Webb2014}. By considering 2 cases, terrestrial Mg isotopic abundances, and the extreme of 100\% $^{24}$Mg, the study in \cite{Webb2024mystery} quantified (by comparing two absorption systems in close redshift proximity) the systematic effect on $\daa$. The results obtained indicate that {\it unless} one can be quite sure of the actual relative Mg isotopic abundances (which will generally {\it not} be the case), it may be prudent to avoid using Mg altogether for varying $\alpha$ measurements. Nevertheless, as Figure \ref{fig:spic_best} illustrates, the MgI 2852.96{\AA} transition was included in the {\sc ai-vpfit} models (the Mg isotopes used here were assumed to be terrestrial abundances). Whilst the $\daa$ values reported in this study may therefore be biased by the inclusion of Mg, the relative changes between $\daa$ values with and without continuum fine tuning (the main point of this paper) are valid.
\end{enumerate}

In addition to searching for new physics via spacetime variations of fundamental constants, another important ELT science driver is the direct measurement of cosmological redshift drift using spectral features in quasar spectra, \cite{Sandage1962} and e.g. \cite{Liske2008}. At higher redshifts, where relatively few continuum segments remain devoid of absorption in the Lyman forest, it is notoriously hard to estimate the local continuum level. For redshift drift measurements, the line position precision required for the measurement of Lyman-$\alpha$ absorption line positions is 2 orders of magnitude greater than for $\daa$ measurements, and it is possible that the impact of variations in the local continuum estimate may dominate other sources of uncertainty. The results presented in this paper motivate a careful investigation of that problem. \\

\section*{Acknowledgements}
We are grateful for access to the OzSTAR supercomputer at the Centre for Astrophysics and Supercomputing at Swinburne University of Technology.

\section*{Data Availability}
The astronomical data used in this study is publicly available from the ESO Archive. The continuum fitting code and {\sc ai-vpfit} result files are available from the authors on request. {\sc vpfit} is publicly available. {\sc ai-vpfit} will be released for general use at a later date (when a user guide has been completed).

\bibliography{contin}
\bibliographystyle{aasjournal}

\widetext

\end{document}